\documentclass{PoS}

\usepackage{epsfig}
\usepackage{latexsym}
\usepackage{amsmath}
\usepackage{amsfonts}
\usepackage{bbm}

\newcommand{\hide}[1]{ }

\newcommand{\rmO}{{\mathcal{O}}}

\newcommand{\eq}{eq.~}

\renewcommand{\vec}[1]{{\bf #1}}
\newcommand{\nwc}{\newcommand}

\newcommand{\tinymsbar}{{\overline{\!\mbox{\tiny\rm{MS}}}}}
\nwc{\nl}  {\newline}
\nwc{\be}  {\begin{equation}}
\nwc{\ee}  {\end{equation}}
\nwc{\fig}{fig.~}
\nwc{\figs}{figs.~}
\nwc{\bmu} {\bar{\mu}}
\nwc{\ba}  {\begin{eqnarray*}}
\nwc{\ea}  {\end{eqnarray*}}
\nwc{\bc}  {\begin{center}}
\nwc{\ec}  {\end{center}}
\nwc{\bi}  {\begin{itemize}}
\nwc{\ei}  {\end{itemize}}
\nwc{\nn}  {\nonumber\\}
\nwc{\Tr}  {\mathop{\rm Tr}}
\nwc{\re}  {\mathop{\rm Re}}
\nwc{\im}  {\mathop{\rm Im}}
\nwc{\Hc}  {\mathop{\rm H.c.}}
\nwc{\la}[1]{\label{#1}}
\nwc{\rmi}[1]{{\! \mbox{\scriptsize #1}}}
\nwc{\nr}[1]{(\ref{#1})}
\nwc{\fr}[2]{{\frac{#1}{#2}}}
\nwc{\msbar}{\overline{\mbox{\rm MS}}}
\nwc{\lambdamsbar}{\Lambda_{\overline{\rm MS}}}
\nwc{\dr}{{4d\to3d}}
\newcommand{\Nf}{N_{\rm f}}
\newcommand{\Nc}{N_{\rm c}}
\newcommand{\Tc}{T_{\rm c}}

\newcommand{\rmii}[1]{{\!\!\mbox{\tiny\rm{#1}}}}

\newcommand{\deltabar}{\,\raise-0.02em\hbox{$\bar{}$}\hspace*{-1.2mm}{\delta}}

\def\lsi{\raise0.3ex\hbox{$<$\kern-0.75em\raise-1.1ex\hbox{$\sim$}}}
\def\gsi{\raise0.3ex\hbox{$>$\kern-0.75em\raise-1.1ex\hbox{$\sim$}}}
\newcommand{\lsim}{\mathop{\lsi}}
\newcommand{\gsim}{\mathop{\gsi}}

\title{Finite-temperature QCD}

\ShortTitle{Finite-temperature QCD}

\author{\speaker{M.~Laine}\\
   Faculty of Physics, University of Bielefeld, D-33501 Bielefeld, Germany\\
        E-mail: \email{laine@physik.uni-bielefeld.de}}

\abstract{I start by discussing recent ideas concerning three 
different heavy quark related observables in finite-temperature QCD.
Subsequently selected studies related to light quarks and gluons
are reviewed, with a focus on thermodynamic observables, 
screening masses, and sum rules.}

\FullConference{The XXVII International Symposium on Lattice Field Theory - LAT2009\\
		 July 26-31 2009\\
		 Peking University, Beijing, China}

\begin{document}

%
\section{Introduction}

The finite-temperature plenary talks at the yearly Lattice conferences
have traditionally been overviews of results obtained during
the past year or so~\cite{lat2006}--\cite{lat2008}. 
This time, using an ``outsider'' status as an 
excuse, I would like to depart from the tradition 
and start with a somewhat more ``active'' approach, 
outlining a few newer ideas with the hope that this 
may help to inspire {\em future work} (sec.~\ref{se:heavy}).
Subsequently, however, I return to time-honoured practices and 
summarize results that caught my attention 
during recent months (sec.~\ref{se:light}).   

More precisely, the basic novelty that I would like 
to elaborate on is the slight 
paradigm shift that has been taking place concerning
the role that heavy quarks (charm and bottom quarks)
may play in hot QCD (with a temperature
$T\sim 150 - 500$~MeV). Given that the masses of the heavy quarks are 
much above the temperature, by up to an order of magnitude, it was 
long thought that they would be relatively ``inert'' and play little
role in this temperature range. 
It has been one of the remarkable empirical discoveries of
the Relativistic Heavy Ion Collider (RHIC) at Brookhaven, though, 
that even within the short lifetime of the thermal system the heavy 
quarks do appear to experience significant interactions 
with it. 
In fact, this is one of the reasons why the medium 
generated in heavy ion collisions
is nowadays conceived to be a ``strongly coupled'' one.

Before embarking on a specific discussion on this topic, 
let me try to place my presentation in a wider context. 
Indeed, high-temperature QCD, or pure SU($\Nc$) 
gauge theory, can be pursued to many different goals. These theories offer, 
for instance, a tractable theoretical limit in which to study various 
aspects of confinement and chiral symmetry breaking, and  
several recent papers as well as 
parallel and poster contributions were formulated in this
spirit (see, e.g., refs.~\cite{conf_chi}). On the phenomenological 
side, the original motivation for considering hot QCD was the 
possible role that it may play in Early Universe cosmology, and 
indeed the relic density of certain dark matter candidates is sensitive 
to the QCD equation-of-state (see, e.g., refs.~\cite{cosmo}).
Currently the most pressing issue of the field is, however, to offer
QCD-based non-perturbative predictions for the observables that 
play a role in on-going and future heavy ion collision experiments, 
so this will be the focus of the present talk.  

%
\section{Heavy quarks at high temperature}
\la{se:heavy}

Traditionally it was assumed, based on leading-order weak-coupling
computations, that charm quarks would not have time to thermalize 
in heavy ion collisions, neither ``kinetically'' nor ``chemically''.
Kinetic thermalization means that the momenta of the heavy quarks 
be distributed thermally, i.e.\ that the average momentum vanish
in the rest frame of the thermal system and the average 
momentum-squared be proportional to the temperature. Chemical 
thermalization means that the number density (or, more properly, 
the entropy density) associated with the heavy quarks be as 
large as thermal field theory predicts.  

A kind of a paradigm shift, associated 
with the concept of a "strongly coupled quark-gluon plasma", has however  
been taking place during the last few years. 
Indeed, experimental observations (to be reviewed below) 
concerning the "quenching" of heavy quark jets can be 
interpreted as indirect evidence for their kinetic thermalization; that 
is, heavy quarks interact more strongly than originally expected. Whether 
a chemical equilibration also takes place is not quite as obvious; 
changes in number density are accompanied by an additional Boltzmann
factor, $\exp(-2 M/T) \ll 1$, where $M$ is a heavy quark mass and the 
factor two accounts for the fact that quarks and antiquarks come in pairs. 
On the other hand, the initial state already contains a distribution of
heavy quarks and antiquarks originating from the very first hard scatterings, 
and this distribution may happen to be of the right order 
of magnitude; if so, only kinetic thermalization is required in order to 
bring them to full equilibrium.  In any case it is now a challenge for 
theorists to, first of all, understand quantitatively the rapid kinetic 
thermalization of heavy quarks, and second, 
just in case, to probe how big an effect
chemically thermalized charm quarks would have in the hydrodynamic 
modelling of heavy ion collision experiments. 

In order to organize the corresponding discussion, I start by considering
the case of ``0 valence'' heavy quarks; by this I refer to the effect of 
(chemically thermalized) heavy sea quarks on the equation of state. 
This case can also be identified as QCD with $\Nf = 2+1+1$ flavours. 
I then proceed to the ``1 valence'' sector, considering heavy quark jets 
and their kinetic thermalization; and end with 
``$1+\bar{1}$ valence'' heavy quarks, meaning heavy quarkonium.
As we will see, there is a perspective for progress on all these fronts.

Before proceeding, I would like to briefly ponder the question of when it is 
precisely that some quark is ``heavy'', compared with the temperature $T$.
To this effect, recall the form of the 
free quark propagator in continuum: 
\be
 \bigl\langle \psi (P)
 {\bar\psi} (Q) \bigr\rangle_0 
  = 
 \deltabar(P - Q)
 \frac{- i \slash \!\!\!\! P \, + M }{P^2 + M^2}
 \;, 
 \quad
 P = (\omega_n,\vec{p}) 
 \;,  \la{psiprop} 
\ee
with $\omega_n = \pm \pi T$, $\pm 3\pi T$, $\ldots\;$. 
We see that the relevant comparison is something 
like $M\leftrightarrow \pi T$, but a question remains whether,
once interactions are taken into account, we should insert the 
$\msbar$ mass, say $M_c^{\raise-0.2em\hbox{$\,\tinymsbar$}}
(\mbox{3~GeV}) \approx 1$~GeV; the pole mass, say  
$M_c^{\,\rmi{pole}} \sim (1.5-2.0)~\mbox{GeV}$; or something else. 
There is probably no unique answer to this question; rather, 
the answer depends on the  observable. In any case, 
the message to take home is that in principle
charm quarks could be ``light'' as soon as 
$T > 1$~GeV/$\pi \sim 300$~MeV, 
or ``heavy'' as long as $T < 2$~GeV/$\pi \sim 600$~MeV.

%
\subsection{ ``0 valence'' --- charm quark effect on $(e-3p)/T^4$}

It was suggested a few years ago, based on a next-to-leading
order weak-coupling analysis, that charm quarks may have a significant
effect on various thermodynamic observables at surprisingly low
temperatures~\cite{pheneos}. The issue then is whether lattice
studies could consolidate this suggestion. 

As far as the lattice goes, it is appropriate to point out that 
although we conceptually now consider QCD with $\Nf = 2+1+1$ flavours, with 
charm quarks in the thermal sea, the lattice estimate of their effect can 
in practice be reduced to the measurement of certain condensates, which
can then be evaluated within the $2+1$ flavour theory. In this approach 
the charm quarks are only partly dynamical. Nevertheless, the order of
magnitude of their effect should still come out right. 

More concretely, focussing on the trace of the energy-momentum tensor, 
i.e.\ $(e-3p)/T^4$, the heavy quark contribution can be 
reduced to~\cite{Cheng:2007wu} 
\be
 \Delta\Bigl( \frac{e-3p}{T^4} \Bigr)  =
 Z_1 \times (\langle\bar\psi \psi \rangle_T 
 - \langle\bar\psi \psi\rangle_0)
 \;, 
\ee
where $Z_1 \propto \partial M_\rmi{bare}/\partial \beta_{\,\rmi{L}}$
is a coefficient function that needs to be determined along the lines of 
constant physics ($\beta_{\,\rmi{L}}$ 
refers to the lattice $\beta$-coefficient). 
Recent measurements, from asqtad fermions with $N_\tau = 6$ and bare mass 
ratios $M_\rmi{bare}/{m_s} \sim {m_s}/{m_{u}} \sim 10$~\cite{Levkova}, 
are compared with the corresponding curve
from perturbation theory in \fig\ref{0valence}. 

%
\begin{figure}[t]

\centerline{%
 \raise8mm\hbox{\epsfxsize=7.0cm\epsfbox{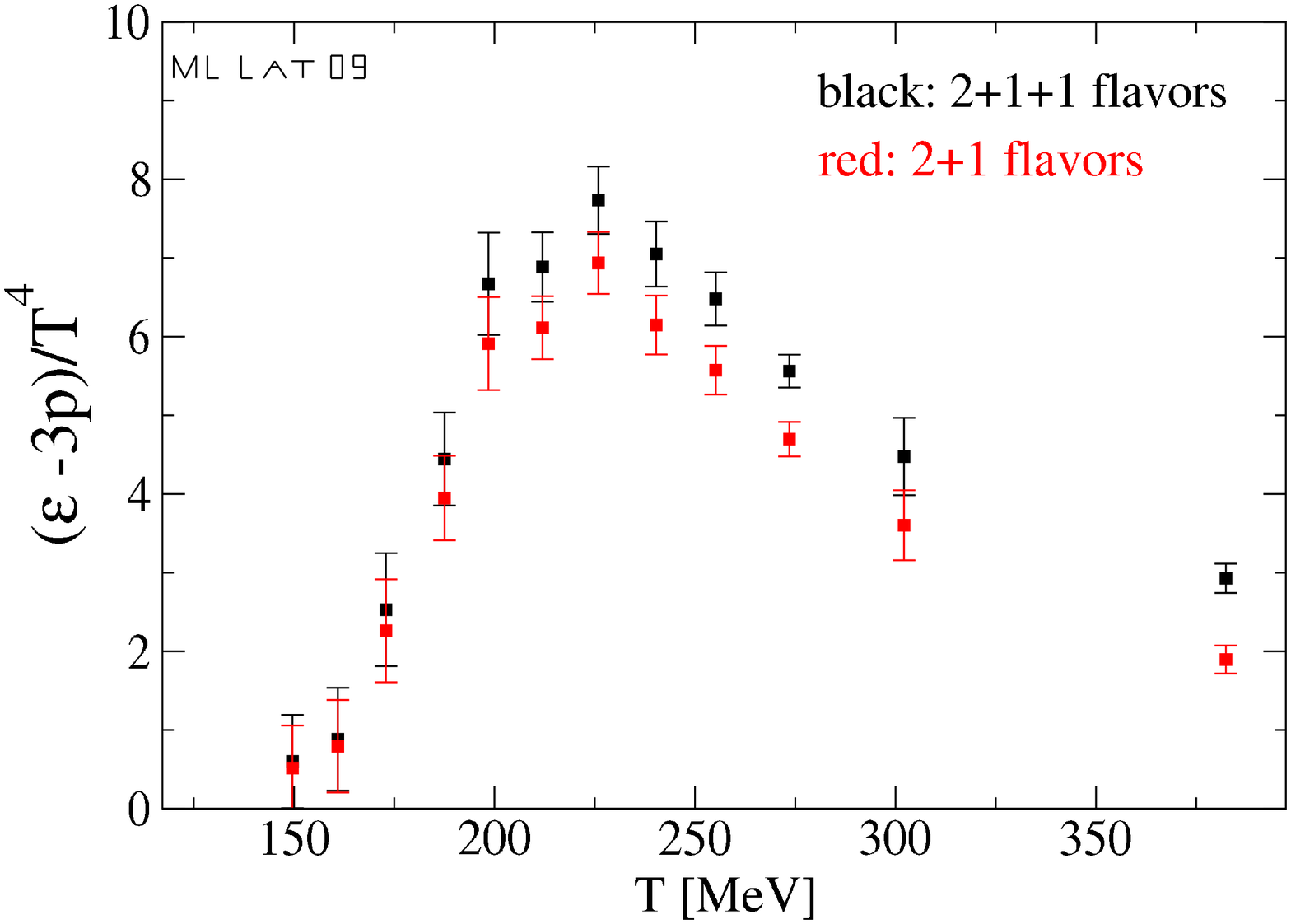}}%
 \hspace*{0.5cm}%
 \epsfxsize=7.0cm\epsfbox{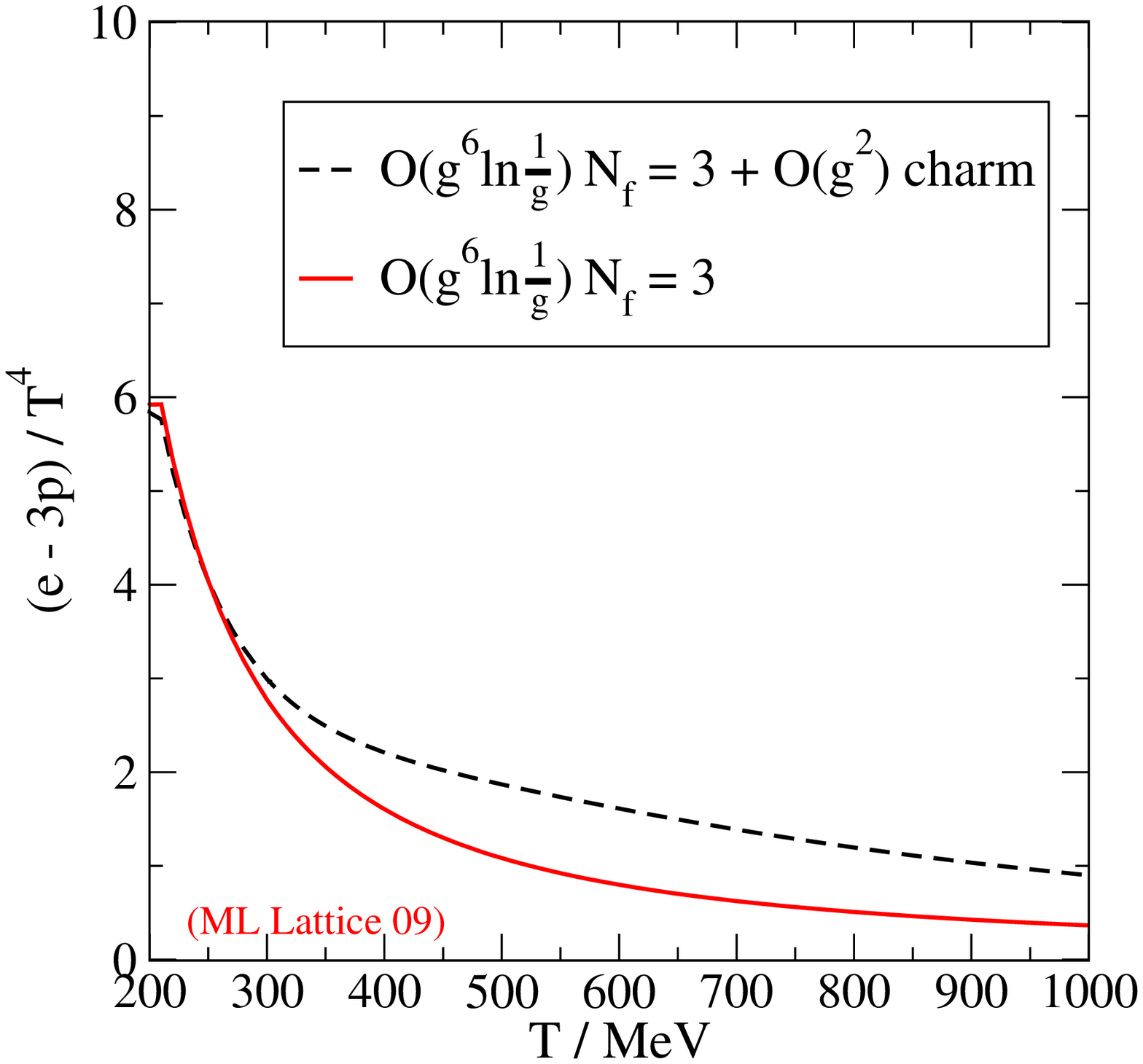} 
}

\caption[a]{\small 
The charm quark contribution to the trace anomaly according
to recent lattice simulations (left)~\cite{Levkova}, as well as 
resummed perturbation theory (right)~\cite{pheneos}. } 
\la{0valence}
\end{figure}
%

We observe that while the effect of 
charm quarks is very small at $T\lsim 250$~MeV, their relative
importance in the trace anomaly increases rapidly with temperature, 
reaching $\gsim 50$\% already at $T\gsim 400$~MeV. (The relative
effect is somewhat smaller in quantities like the pressure or
energy density, where no subtraction is carried out.)

To summarize, if chemical equilibration takes place, 
then charm quarks might affect the initial stages 
of hydrodynamics in future heavy ion collisions at the LHC, where 
higher temperatures may be reached than at the RHIC. In any case, the
charm quarks do play a significant role in the equation of state 
relevant for cosmology, in which environment chemical equilibrium 
is guaranteed to be reached for all strongly interacting particles. 

%
\subsection{``1 valence'' --- heavy quark jets}
\la{ss:1valence}

In any high energy collision, a number of heavy quarks (and antiquarks)
are produced in an initial hard process, as can be illustrated by the 
following Feynman diagram:  


\vspace*{0.3cm}

\centerline{
\begin{minipage}[c]{3.5cm}
\epsfxsize=3.6cm\epsfbox{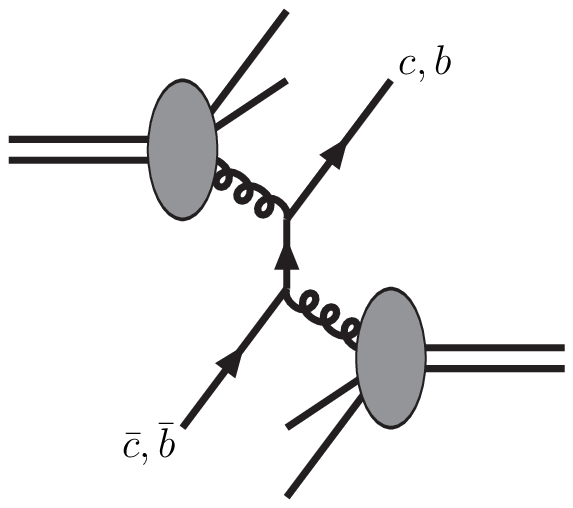} 
\end{minipage}
}

\vspace*{0.3cm}


\noindent
Assuming a suitable factorized framework and making use of previous
experience gathered in $p + \bar{p}$ or $p + p$ collisions, it is believed
that the corresponding 
production cross section is relatively well understood also 
for collisions involving heavy nuclei, such as {\it d + Au} and
{\it Au + Au}~\cite{cnv}. Subsequent to their production, the 
heavy quarks decay, often semi-leptonically as $c\to \ell\nu X$. 
The leptons $\ell$ can be observed, and the outcome
can then be compared with the theoretical prediction. 
 
It turns out that while results from {\it d + Au} collisions indeed
conform with theoretical expectations, those from {\it Au + Au} collisions
appear not to do so; rather, less leptons $\ell$ are observed than expected.  
The results are illustrated in \fig\ref{exp} in terms of the 
so-called ``nuclear modi\-fication factor'', $R_{AA}$, 
which is significantly below unity for {\it Au + Au} collisions~\cite{star}; 
we can say that heavy quark jets get ``quenched''.
Another relevant observable is the so-called elliptic flow, and the indication
is that heavy quarks do participate in this hydrodynamic behaviour~\cite{phenix}. 
Both observations point towards the interpretation that, due to multiple
scatterings with other particles, the heavy quarks
slow down with respect to the thermal medium, 
and then flow together with it. I refer to this phenomenon 
as kinetic thermalization. In fact, the heavy quarks behave much like 
heavy particles in classical non-relativistic Brownian motion, 
and many of the same concepts can be argued to apply~\cite{cst}.

\begin{figure}[tb]


\centerline{
\epsfxsize=8.0cm\epsfbox{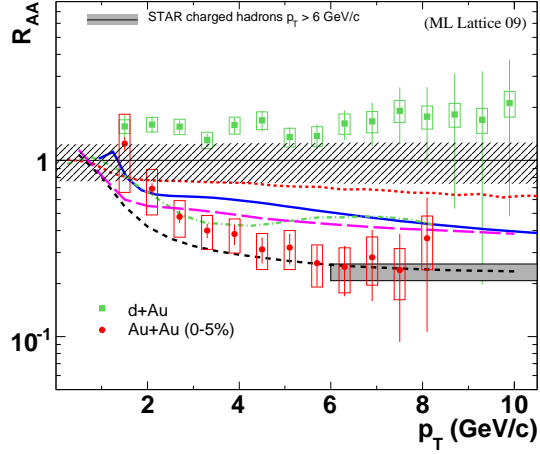} 
}
\caption[a]{\small 
Experimental results from ref.~\cite{star} (see also ref.~\cite{phenix}), 
indicating that heavy quark jets propagate as expected in {\it d + Au}
collisions, but get quenched in {\it Au + Au} collisions.  } 
\la{exp}
\end{figure}

Let us denote the rate at which heavy quarks, assumed already to 
be close to rest, slow down, by $\eta_D$; this thermalization rate
is often also called the ``drag coefficient''. Inspired by the analogy 
with Brownian motion and its classical description through Langevin
dynamics, $\eta_D$ can be fluctuation-dissipation-related 
to another coefficient, $\kappa$, characterizing the autocorrelation
of the force that acts on the heavy quarks~\cite{cst,eucl}:  
\be
 \eta_D  =   \frac{\kappa}{2 M_\rmi{kin} T}
 \left(1 + 
 O\!\left(\frac{\alpha_s^{3/2}T}{M_\rmi{kin}}\right)\right) \;, \quad
 \kappa  =   \lim_{\omega\to 0} 
 \frac{2 T \rho_E(\omega)}{\omega} 
 \;, \la{master}
\ee 
where $M_\rmi{kin}$ refers to a particular heavy quark mass 
definition, related to the pole mass at zero temperature. 
The function $\rho_E$ is the spectral function corresponding 
to the Euclidean correlator 
(assuming $T > \Tc$ or $\Nf > 0$)~\cite{eucl} 
\be
 G_E(\tau) = - \fr13 \sum_{i=1}^3 
 \frac{
  \Bigl\langle
   \re\Tr \Bigl[
      U_{\beta;\tau} \, gE_i(\tau,\vec{0}) \, U_{\tau;0} \, gE_i(0,\vec{0})
   \Bigr] 
  \Bigr\rangle
 }{
 \Bigl\langle
   \re\Tr [U_{\beta;0}] 
 \Bigr\rangle
 }
 \;, \la{GE_final}
\ee
where $g E_i  \equiv i [D_0,D_i]$ is the colour-electric field 
(shown here in continuum notation), 
and $U_{\tau_\rmii{\;b},\tau_\rmii{\;a}}$ is a Wilson line 
in the Euclidean time direction, from $\tau_\rmii{\;a}$ to $\tau_\rmii{\;b}$.
The numerator can be illustrated as 


\vspace*{0.5cm}

\centerline{
\hspace*{0.5cm}%
\begin{minipage}[c]{2.0cm}
\epsfxsize=2.0cm\epsfbox{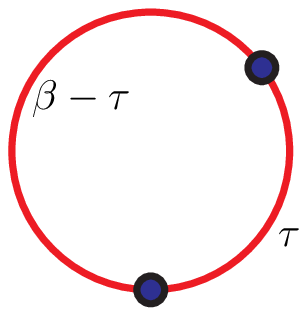} 
\end{minipage}~~~
}

\vspace*{0.3cm}

%

\noindent
where the circle represents the Polyakov loop around the Euclidean
time direction and the blobs denote electric field insertions.
As far as I know no lattice measurements of this correlator have been 
published yet, although they should not be overwhelmingly demanding. 

\begin{figure}[tb]

\centerline{%
\epsfxsize=7.5cm\epsfbox{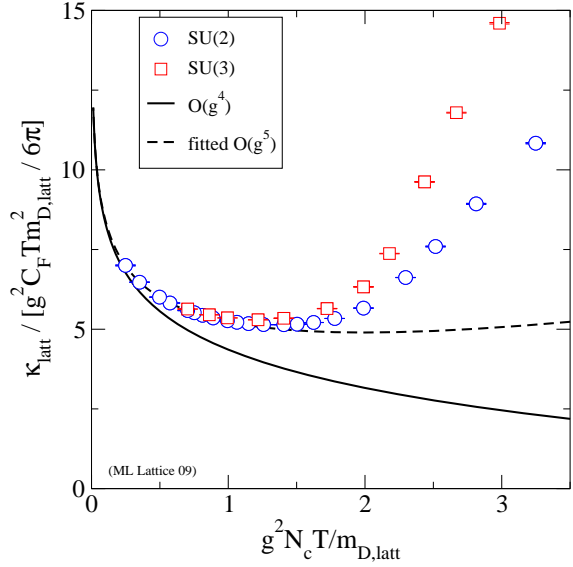} 
}

\caption[a]{\small 
Results for the force-force transport coefficient $\kappa$
within classical lattice gauge theory, plotted as a function
of the perturbative expansion parameter
$g^2 \Nc T / m_{\,\rmi{D,latt}}$ (from ref.~\cite{mink}).
The non-perturbative data can exceed 
the leading-order and even the next-to-leading order 
results by a significant amount. } 
\la{fig:mink}
\end{figure}

Even though no actual measurements are available, another test
has been carried out. Indeed, the electric field correlator can also
be addressed within ``classical lattice gauge theory'', which serves
as kind of an effective low-energy description of infrared phenomena
at finite temperatures~\cite{clgt}. It turns out that classical 
lattice gauge theory is more sensitive to ultraviolet physics than 
would be expected from a proper effective field theory framework; 
nevertheless, the result is interesting if plotted in terms of 
a quantity that does have a direct analogue in QCD, the so-called Debye mass 
parameter, $m_\rmi{D,latt}^2 \sim g^2 T/a$. A result is shown 
in \fig\ref{fig:mink}, including comparisons with a leading
order perturbative result and a fitted next-to-leading order 
behaviour (the next-to-leading order result has been computed 
analytically in continuum QCD~\cite{sch}, but not in classical 
lattice gauge theory, where only its parametric form is known).

It can now be observed that if we insert the estimate
$g^2T \sim m_\rmi{D,latt}$, which is known to be a reasonable 
one on the QCD side, then we are in a regime where 
the next-to-leading order correction is of order 100\%, 
yet the non-perturbative result is larger still.  
Excitingly, such a significant increase appears to be 
more or less on top of what is phenomenologically needed in order
to explain the fast thermalization (quenching) of heavy quark jets
(see, e.g., ref.~\cite{th} and references therein).

\begin{figure}[tb]


\centerline{
\epsfxsize=7.0cm\epsfbox{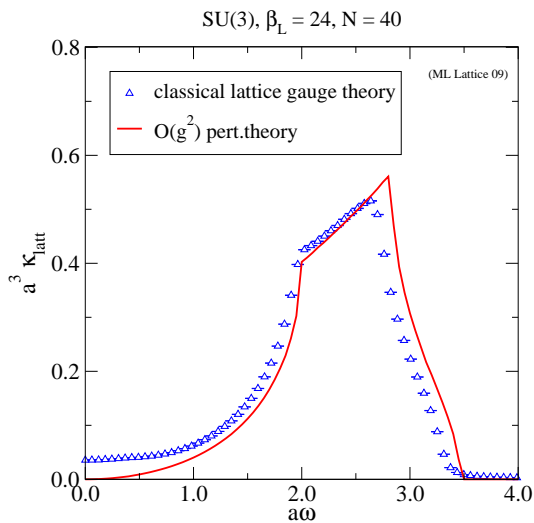} 
}

\caption[a]{\small 
The combination ${2 T \rho_E(\omega)}/{\omega}$ from \eq\nr{master}, 
in lattice units, according to classical lattice
gauge theory (from ref.~\cite{mink}). The intercept at $\omega = 0$
yields the result plotted in \fig\ref{fig:mink} in different units.} 
\la{fig:rhoE}
\end{figure}

Classical lattice gauge theory makes another 
prediction as well. In \fig\ref{fig:rhoE}, the frequency-dependent
function from which the intercept is to be taken according to 
\eq\nr{master}, is shown. The basic observation is that this 
function is {\em flat} at small frequencies; it has  
{\em no transport peak}, unlike spectral functions related to 
conserved currents. A similar flat behaviour for the spectral
function related to the electric field correlator has also been 
observed in a very different theory, strongly coupled
${\mathcal N} = 4$ Super-Yang-Mills theory in the 
large-$\Nc$ limit, handled through its AdS/CFT dual~\cite{sg}.

All this suggests that the electric field correlator might be 
more amenable to analytic continuation than current--current correlators
from which viscosities and conductivities have been extracted
previously~\cite{recent}. To summarize, I can only solicit 
numerical tests of \eq\nr{GE_final}, and hope that the 
outcome might yield a large coefficient $\kappa$, differing
from the perturbative one by as much as an order of magnitude.  
With the advent of LHC, which should produce data on bottom quark
jets as well, containing a different $M_{\,\rmi{kin}}$
in \eq\nr{master} than in the charm quark case, 
our understanding of heavy quark jets
within a hot medium could then be quantitatively tested. 

%
\subsection{``$1+\bar{1}$ valence'' --- heavy quarkonium}
\la{ss:qbq}

As a last example of heavy quark related observables, I briefly
summarize recent news from heavy quarkonium physics. Like heavy 
quarks, heavy quarkonium can originally be generated in a non-thermal
hard scattering, or through a thermal fluctuation. The latter process is 
reminiscent of those contributing to the chemical thermalization 
of single heavy quarks, and therefore probably too slow to take 
place effectively; nevertheless, surprises cannot be excluded. 
After having formed one way or the other, heavy quarkonium propagates
through the thermal medium, whereby its properties get modified; 
therefore the quarkonium peak observed in the dilepton rate~\cite{dilepton} 
may change in magnitude, shape, or position, depending on the temperature 
that is reached in the collision~\cite{ms}. Some relevant Feynman 
diagrams are illustrated in \fig\ref{fig:diags}.

\begin{figure}[tb]


\centerline{%
\epsfxsize=6.0cm\epsfbox{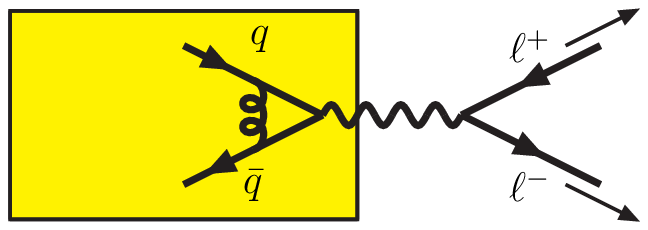}~~~%
\hspace*{1cm}%
\epsfxsize=6.0cm\epsfbox{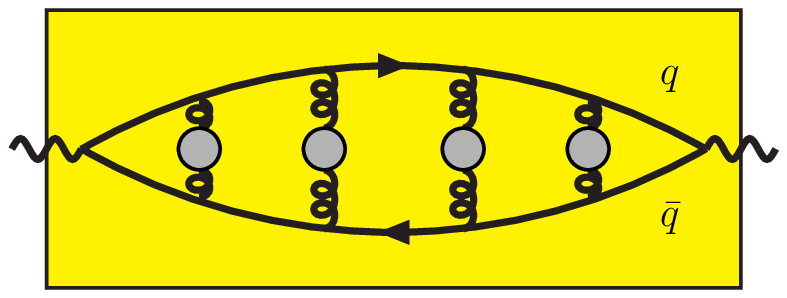}~~~%
}


\caption[a]{\small 
Left: an amplitude corresponding to the production 
of a dilepton pair from a thermalized medium. 
Right: squaring the amplitude,  
the dilepton production rate is seen to be  
proportional to the two-point correlation function 
of the electromagnetic current, with significant near-threshold 
corrections coming from medium-modified Coulomb exchange. 
} 
\la{fig:diags}
\end{figure}

Now, like always in quarkonium physics, it appears reasonable
to try and address the thermal modification of quarkonium properties 
through a potential model, thereby resumming corrections
from graphs of the type in \fig\ref{fig:diags}(right). At finite
temperatures, however, this is complicated by the 
multitude of different potentials that can in principle be defined, 
and intuitive arguments alone cannot decide which of them is 
the correct one. What is needed is rather a derivation of the 
relevant effective framework from QCD; at least
within perturbation theory this can 
indeed be achieved~\cite{static}, resulting in a potential-model type recipe
for computing the spectral function~\cite{peskin}, with a definite 
(in general complex) potential appearing 
as a ``matching coefficient''. 
 
In order to appreciate the intricacies of the issue, 
it is important to realize that,
on a Euclidean lattice, the time extent $\beta=1/T$ is in some 
sense always ``small''; more quantitatively, it can 
be argued that quarkonium melts at a temperature where parametrically 
$\beta <  1/\alpha_s M_\rmi{kin}$~\cite{peskin}.
In contrast, the Minkowskian time scale $t$ 
corresponding to the Coulombian binding energy of heavy quarkonium  
is ``large'', $t \sim 1/\alpha_s^2 M_\rmi{kin}$. 
So, in the heavy quark limit where the effective 
$\alpha_s$ is small, we see that $t \gg \beta$, and it is more 
or less clear that the potential relevant for discussing quarkonium
binding and dissociation at finite temperatures 
should involve some sort of an analytic continuation.

In this conference, a very interesting suggestion for a non-perturbative
definition of a real-time static potential in this spirit 
was put forward~\cite{akr}. 
Motivated by the perturbative definition, the idea is to first measure 
a Wilson loop as a function of a Euclidean time coordinate, $\tau$; 
this can be illustrated as 


\vspace*{0.3cm}

\centerline{
\epsfxsize=3.0cm\epsfbox{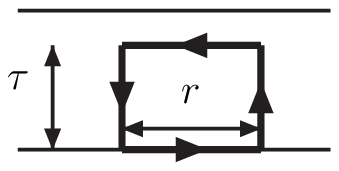} 
}

\vspace*{0.1cm}


\noindent
which observable we denote by 
$
   C_E(\tau,r)  \equiv   
  \langle \Tr[W_E(\tau,r)] \rangle
$.
Supposing that an analytic continuation can be carried out, 
$\tau\to i t$, a real-time potential
could then be extracted from 
\ba
  i \partial_t C_{E}(it,r) & \equiv & V_{>}(t,r) C_{E}(it,r)
 \;.
\ea
Its static limit corresponds to $V_>(\infty,r)$. 

It is appropriate to point out that the 
horizontal Wilson lines appearing in the definition
of $C_E(\tau,r)$ are non-unique, as usual; however, 
to the extent that we can compensate for the specific 
choice by a normalization factor, $Z^{1/2}_s(r)$, so that 
the correlator reads $C_E(\tau,r) = Z_s(r) \tilde C_E(\tau,r)$, 
we see that $Z_s(r)$ drops out from the definition of $V_{>}(t,r)$. 
Such a normalization factor is indeed characteristic of 
the effective field theory framework that can be used 
for addressing the properties of heavy quarkonium~\cite{nb2}.

Now, according to ref.~\cite{akr}, the static limit
$V_>(\infty,r)$ can indeed be extracted through 
a spectral analysis of the Euclidean correlator. More precisely, 
the data appear to indicate 
the presence of a spectral peak, whose position signals
the average energy of the quark-antiquark system, $\re V_{>}(\infty,r)$.
The peak should have a finite width as well, $\im V_{>}(\infty,r)$, 
being a signal of a Coulomb scattering/Landau damping
induced ``decoherence'' of the quark-antiquark state~\cite{static}, 
caused by collisions with the particles of the thermal medium. 

To me, the idea of ref.~\cite{akr} seems very interesting, and 
I am looking forward to further developments along these lines. 

%
\section{Light quarks and gluons at high temperature}
\la{se:light}

I now move away from heavy quark related observables and discuss 
a number of recent developments related to light quarks and gluons.

%
\subsection{Basic thermodynamics with $\Nf = 2 + 1$, $\Nc = 3$}
\la{se:basic}

\begin{figure}[tb]


\centerline{%
\epsfxsize=7.0cm\epsfbox{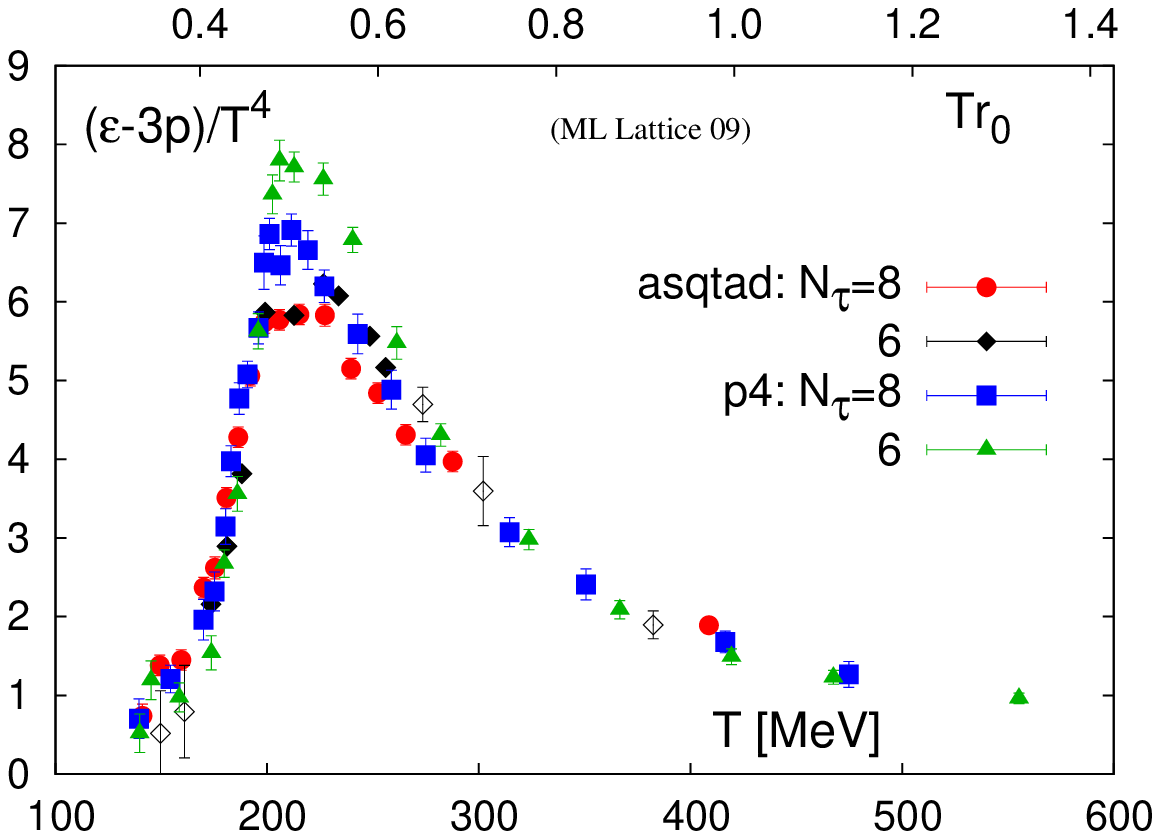}~~~%
\hspace*{1cm}%
\raise4mm\hbox{\epsfxsize=6.0cm\epsfbox{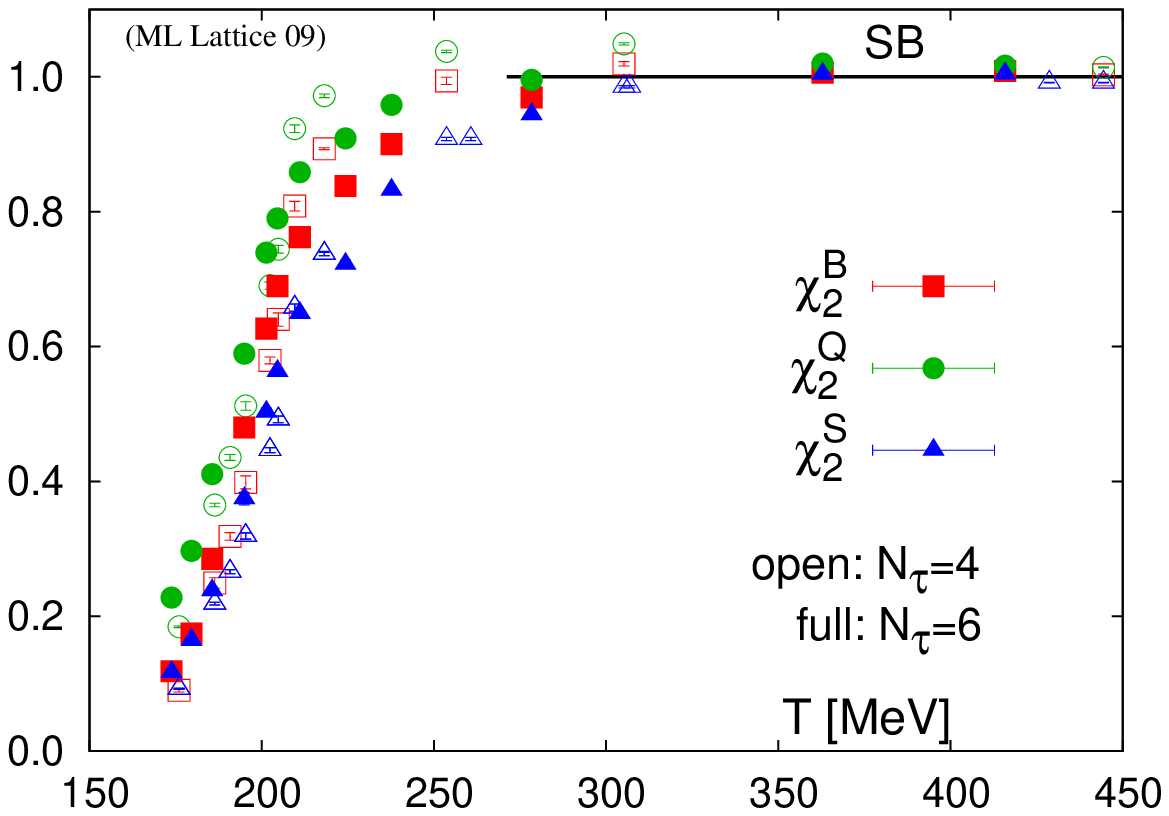}}~~~%
}


\caption[a]{\small 
Left: the trace anomaly $(e-3p)/T^4$ from ref.~\cite{hotQCD}, with two 
families of fermion discretizations. Judging with bare eye, the results 
appear to converge towards a common continuum limit. 
Right: baryon number, electric charge, and strangeness susceptibilities
($\chi_2^B$, $\chi_2^Q$, $\chi_2^S$, respectively), from ref.~\cite{rbc1}.
} 
\la{fig:eos}
\end{figure}

Results for the trace of the energy-momentum tensor
(like in \fig\ref{0valence}) and for various physical 
susceptibilities, from the large-scale simulations in 
refs.~\cite{hotQCD,rbc1}, are shown in \fig\ref{fig:eos}. 
These results are supposed to be ``physical'', i.e.\ for
(almost) realistic quark masses, and the lattice spacing
dependence would appear to be regular as well. 
Improved staggered quarks were used. 

It is fair to say, though, that it is not easy to judge the systematic 
uncertainties that may still be hidden in these results.
For instance, in \fig\ref{fig:bw} a comparison of the strangeness 
susceptibilities, as determined by the RBC-Bielefeld and by
the Budapest-Wuppertal collaborations, is shown~\cite{bw}.
The infamous 30 MeV temperature shift is clearly visible. 

\begin{figure}[tb]

\vspace*{0.3cm}

\centerline{%
\epsfxsize=6.0cm\epsfbox[18 434 290 695]{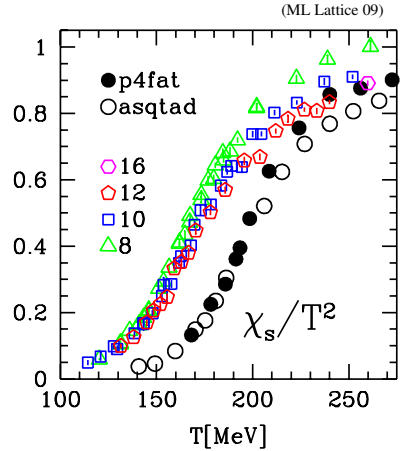}~~~%
}

\vspace*{-0.3cm}

\caption[a]{\small 
The strangeness susceptibilities from different collaborations, 
according to ref.~\cite{bw}.
} 
\la{fig:bw}
\end{figure}

The good news is that the observable
of \fig\ref{fig:bw} is a nice one: it is directly physical, 
being related to the zero component of a conserved current, 
so that no renormalization ambiguities should appear. I encourage
other groups to also primarily carry out comparisons with this quantity, 
rather than with more ultraviolet sensitive ``auxiliary'' observables, like 
the susceptibilities related to the Polyakov loop or the chiral condensate. 

As far as the reason for the discrepancy goes, I am certainly not qualified
to offer any original insight. I do assume that ``trivial'' issues, 
like improper thermalization (see, e.g., ref.~\cite{thermal}), 
have been carefully excluded by all collaborations. 
As a ``user'' I do find it a pity, though, that none of the collaborations
shows finite-volume scaling in their plots, even though the chiral 
limit is not far and the transition is very weakly of the first order; 
indeed \figs\ref{fig:eos}, \ref{fig:bw} refer to a fixed
box size $L$ in units of the temperature, $LT = 3 - 4$.\footnote{%
 Very recently scaling studies in the critical region have 
 been reported in ref.~\cite{Unger}.
 }
On the side
of discretization effects, it would be nice to overlay results
from Wilson-like discretizations on \fig\ref{fig:bw}, particularly 
given the delicate role the chiral symmetry plays close to the transition
point, and fortunately efforts in this direction appear to be 
under way~\cite{MMP,Kanaya,Schierholz}.

%
\subsection{New precision for $\Nf =0$, $\Nc = 3$} 
\la{ss:nf0}

Recently, there have been new studies of the thermodynamics 
of pure SU(3) gauge theory. The great benefit of this simplified theory is 
that systematic errors can be brought better under control; therefore 
the theory offers an excellent test bench both for new lattice ideas, 
and for comparing the lattice data with various continuum computations. 
These two aspects are illustrated in \fig\ref{fig:nf0}. On the left, a test
is shown of the new approach of ref.~\cite{whot}, in which the lattice spacing
is kept fixed and the temperature is varied through changing the number
of points in the Euclidean time direction, $N_\tau$; this is theoretically
more transparent than the standard approach where $N_\tau$ is kept fixed
and temperature is varied through $\beta_{\,\rmi{L}}$, implying a simultaneous
variation of the lattice spacing. On the right, a very precise study 
of the entropy density at low temperatures is shown; the entropy 
density is a convenient observable in that it can be measured without
any subtractions (because 
the entropy density of the vacuum state vanishes), through
\be
 s  =\fr{4}{3T} {Z}_2 \Tr [\vec{B}^2 - \vec{E}^2]
 \;, \la{eq:s}
\ee 
where ${Z}_2$ is a renormalization factor. It can be seen that  
the results are precise enough to allow for a stringent comparison
with the contributions from various glueball spectra. 

\begin{figure}[tb]


\centerline{%
\epsfxsize=6.5cm\epsfbox{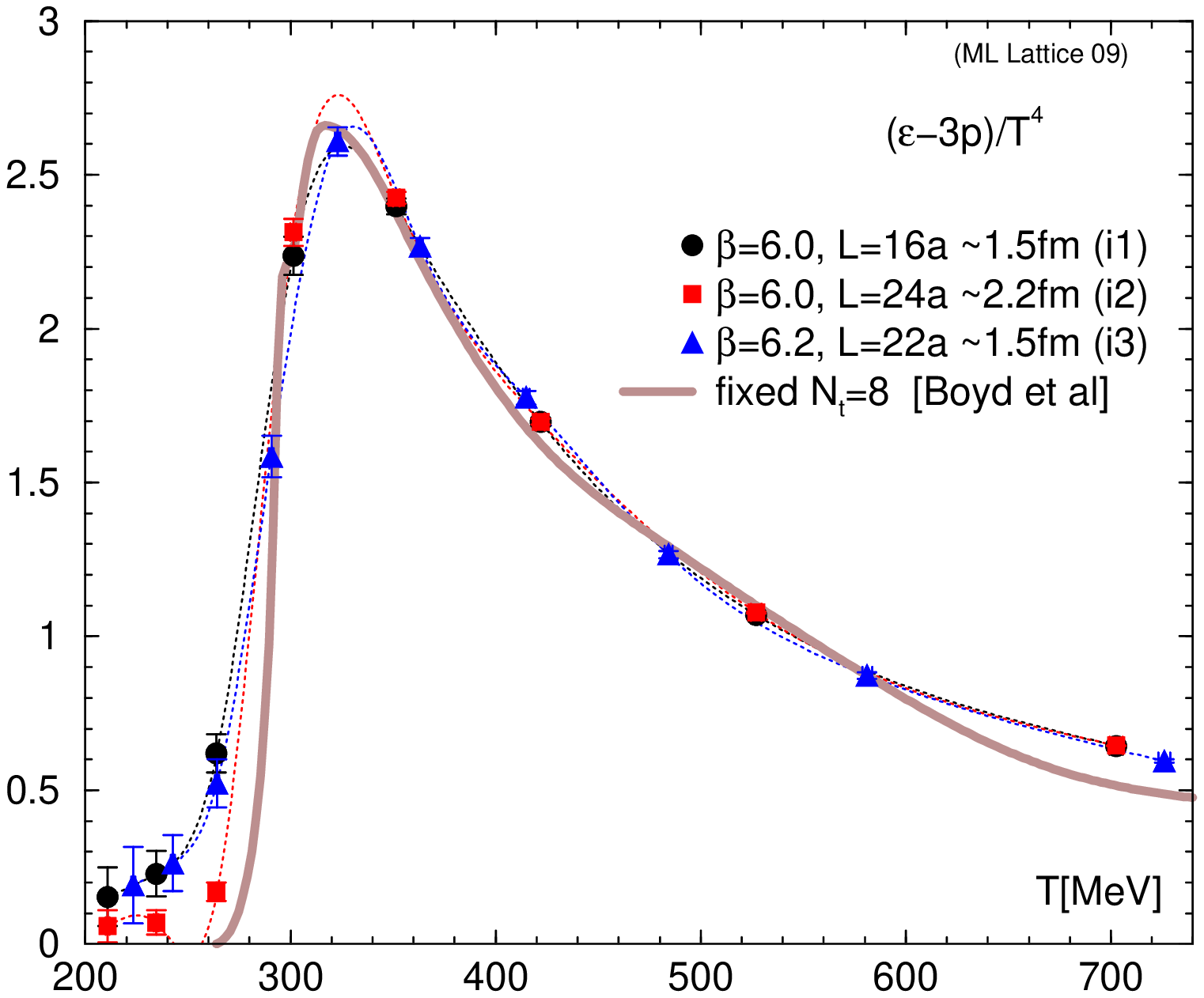}~~~%
\hspace*{0.5cm}%
\raise5.2cm\hbox{\includegraphics[width=5.4 cm,angle=-90]{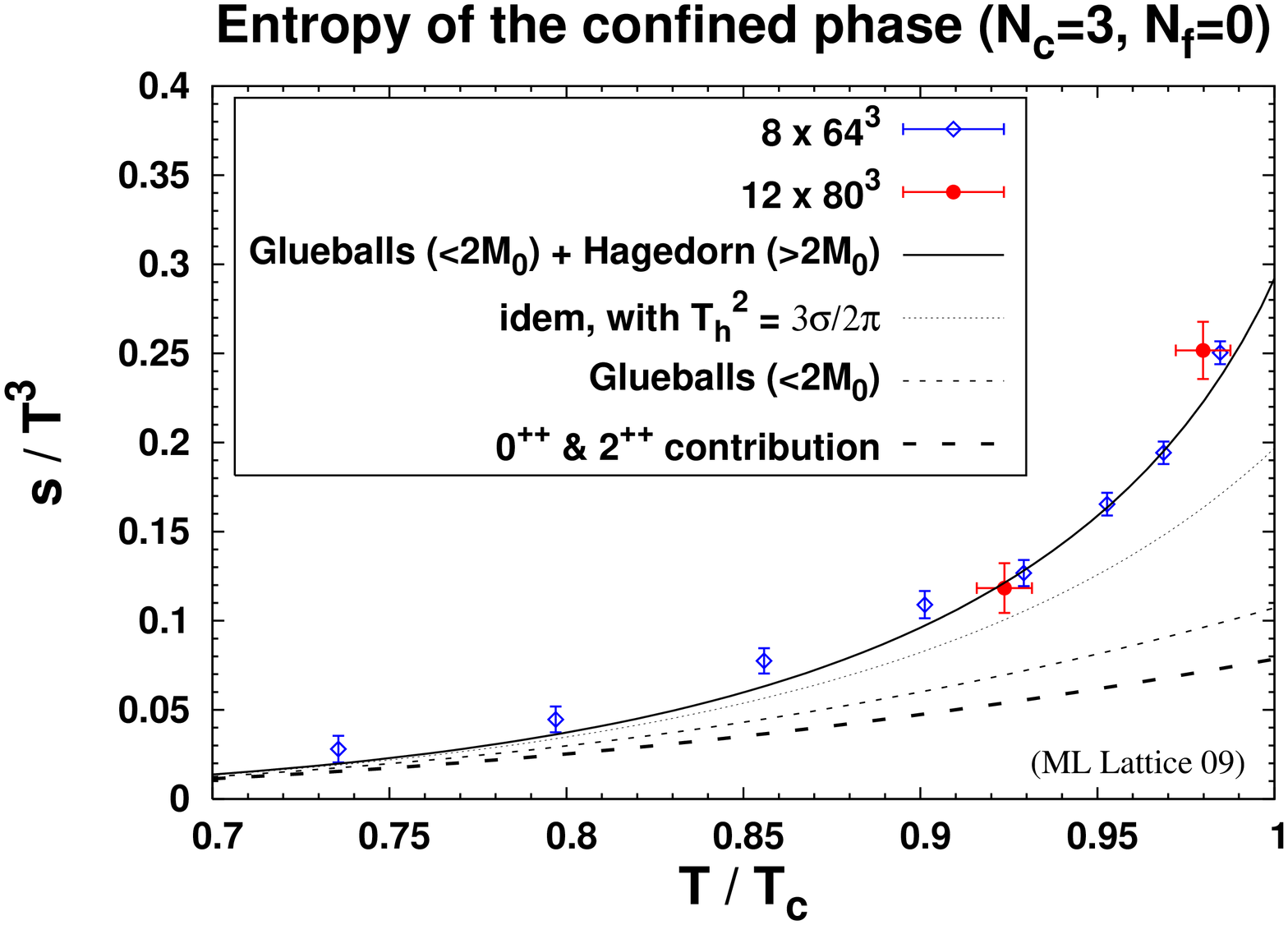}}~~~%
}


\caption[a]{\small 
Left:  the trace anomaly $(e-3p)/T^4$ from ref.~\cite{whot} for $\Nf = 0$, 
in the so-called fixed-scale approach. 
Right:  the entropy density from ref.~\cite{hbm1} for $\Nf = 0$, together 
with a comparison with various glueball resonance descriptions.
} 
\la{fig:nf0}
\end{figure}

In my opinion, this kind of precision relooks at pure gauge 
theory are very welcome, and deserve to be pursued for many other
observables as well. 

%
\subsection{Another look at $\Nf =0$, $\Nc > 3$} 
\la{ss:nc4}

Building on the previous section, it is also interesting to inspect
pure gauge theories with $\Nc > 3$. Recently two groups have 
come up with new results in this spirit (cf.\ \fig\ref{fig:nc}).
In ref.~\cite{dg1}, results have been presented for $\Nc = 3,4,6$, 
with a focus on scale setting and continuum extrapolation ($N_\tau = 6, 8$). 
Another work was presented in ref.~\cite{mp} where, 
following the earlier work in ref.~\cite{bt}, results were 
presented for $\Nc = 3,4,5,6,8$, for a fixed $N_\tau = 5$.

A very intriguing trend can be extracted from the results
of ref.~\cite{bt} (\fig\ref{fig:nc}(right)). Indeed, it appears 
that for large $\Nc$, the functional form of the trace anomaly
becomes much simpler than for $\Nc = 3$: once normalized to 
$\Nc^2 T^4$, the result is basically zero for $T < \Tc$; 
displays a large jump (a first order transition) at $T = \Tc$; 
and decreases then monotonically for $T > \Tc$. This is a very 
simple pattern which suggests, and even calls for, a theoretical 
explanation; in fact, it can perhaps be speculated that
in this limit the high-temperature phase is 
qualitatively purely perturbative.

\begin{figure}[tb]


\centerline{%
\epsfxsize=7.0cm\epsfbox{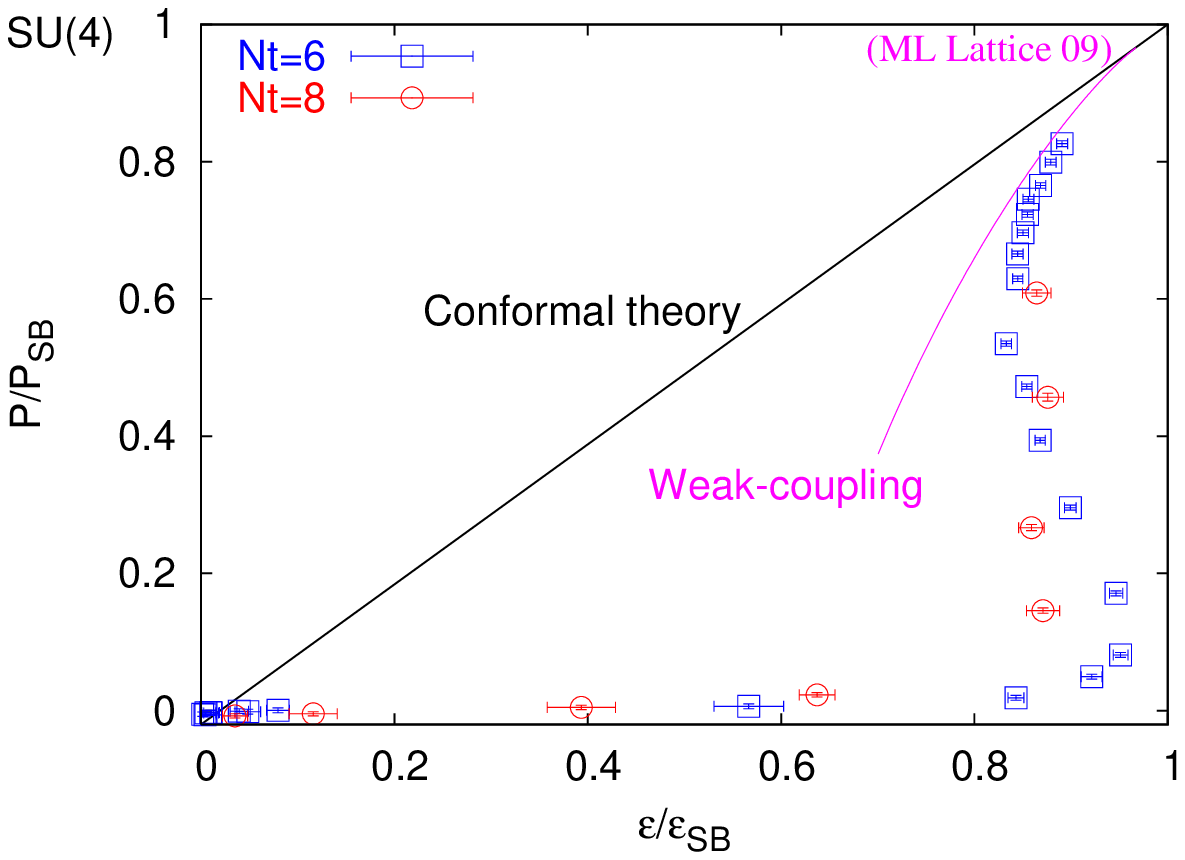}~~~%
\hspace*{0.5cm}%
\raise0.0mm\hbox{\epsfxsize=7.0cm\epsfbox{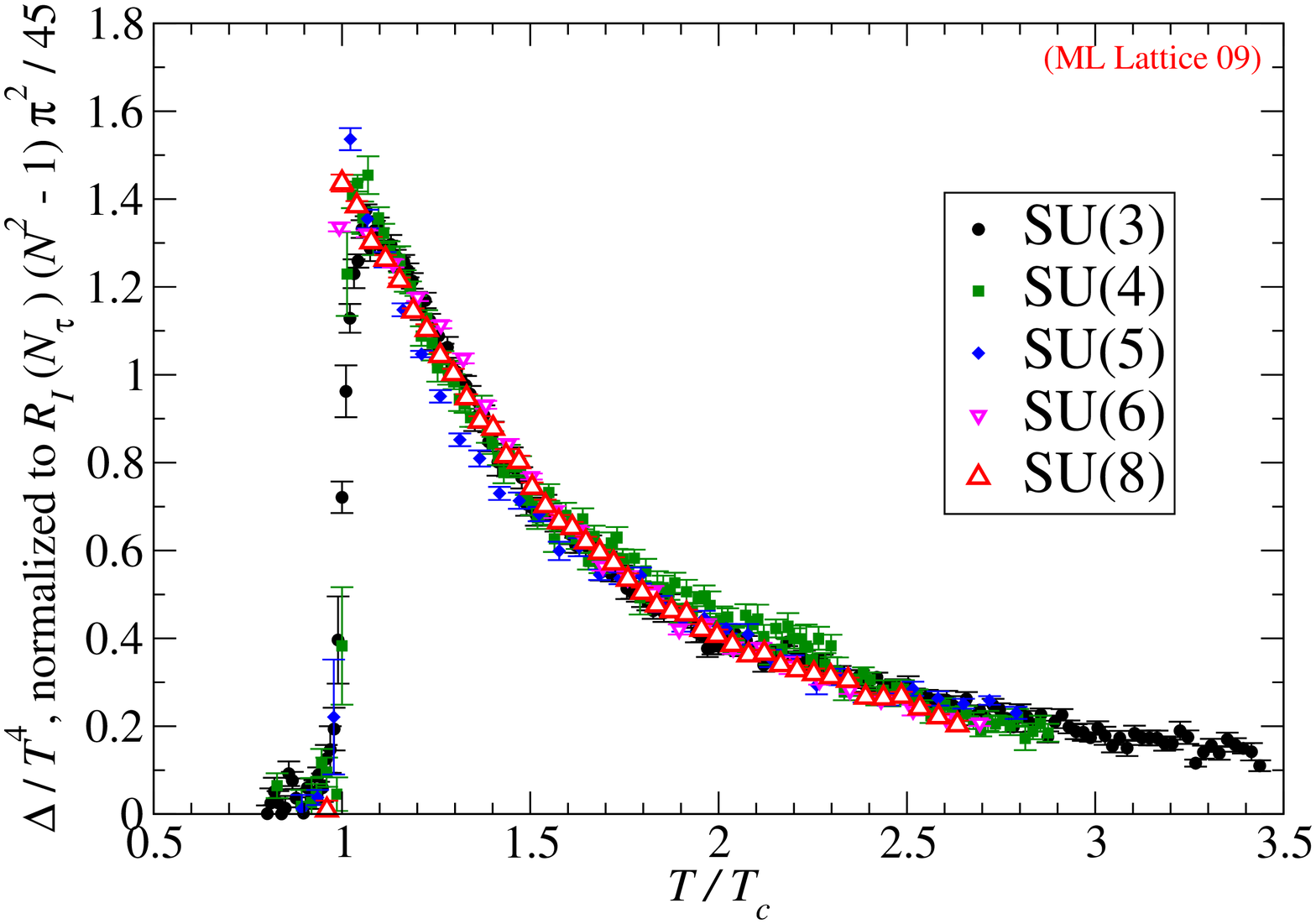}}~~~%
}


\caption[a]{\small 
Left:  pressure versus energy density for SU(4) gauge theory, 
compared with the conformal theory as well as with the weak-coupling
prediction (from ref.~\cite{dg1}).
Right:  the trace anomaly $(e-3p)/T^4$ for various $\Nc$, 
at a fixed $N_\tau =5$ (from ref.~\cite{mp}).
} 
\la{fig:nc}
\end{figure}

A word of caution may be in order, though. In ref.~\cite{wlt}, 
the latent heat (the discontinuity of the energy density or, 
for that matter, of the trace anomaly) was studied at various $\Nc$
for $N_\tau = 5,6,8$, and it was found that in general the results
at $N_\tau = 5$ did not fit well into the continuum extrapolation, 
particularly at large $\Nc$. Given that in ref.~\cite{mp} only $N_\tau = 5$ 
was considered, the results should probably be assigned a generous systematic
error (as indeed elaborated upon in ref.~\cite{mp}); 
in particular, it might be desirable to have additional 
$N_\tau$'s to add confidence to  the $\Nc\to\infty$ limit
(for $\Nc = 4,6$, the results of ref.~\cite{dg1} 
might be helpful in this exercise). 
In any case, I consider this to be a very interesting topic 
and worth further study. 

%
\subsection{Finite-volume effects and screening masses}
\la{ss:finiteV}

In section~\ref{se:basic}, I already alluded to finite-volume effects. 
Recently, an interesting theoretical work appeared~\cite{hbm2} in which
the finite-volume effects in various thermodynamic observables were 
analyzed in some detail. For instance, for the entropy density, 
$s=(p+e)/T$, which can be measured according to \eq\nr{eq:s}, 
the expression 
\be
 s(T,L) = s(T,\infty) - 
 \nu \frac{ e^{-m(T)L}}{2\pi L}
 [1  + \fr32 T \partial_T ] m^2(T) + \rmO\Bigl(e^{-\sqrt{2} m(T) L}\Bigr)
 \la{finiteV}
\ee
was given, with $m(T)$ denoting the lightest screening 
mass and $\nu$ its degeneracy. Similar formulae were also given for
combinations like $e-3p$ and $p$; the ones for $e-3p$  
would play a role in measurements such as those 
in \figs\ref{fig:eos}(left), \ref{fig:nf0}(left), \ref{fig:nc}(right). 
The exponential dependence in \eq\nr{finiteV} is familiar, 
but it is interesting that the pre-exponential factor is also 
completely fixed in terms of known or measurable quantities. 

In my opinion, 
the fact that finite-volume effects are determined by the lightest 
screening mass, as exemplified by \eq\nr{finiteV}, underlines the 
usefulness of measuring the screening masses in the context of 
every lattice study. Indeed, though of no direct use for heavy
ion experimentalists, the screening masses allow theorists 
to learn a lot about the dynamics of the system, 
and also to judge whether systematic errors related to 
finite-volume effects can be under control.

Examples of measurements of screening masses
are shown in \fig\ref{fig:mT}. Among the qualitative
observations that can be made from these plots are the following: 
\bi

\item 
For $T \gsim 1.5 \Tc$, the screening masses coupling dominantly
to gluonic objects (i.e.\ flavour singlets) are the lightest ones. 
That this should happen is one of the predictions of the dimensionally
reduced effective description of high-temperature QCD~\cite{dr}; thus, 
we may assume that dimensional reduction could work qualitatively 
for $T \gsim 1.5 \Tc$ while it probably cannot capture all the relevant 
dynamics for $T \lsim 1.5 \Tc$ if physical $\Nc$ and $\Nf$ are being used. 

\item
It can be seen that {\em cutoff effects} in the mesonic screening mass become 
substantial at {\em high temperature}. It is interesting to compare
this behaviour with that in \fig\ref{fig:bw}; much the same appears
to be the case in the strangeness susceptibility. 

\item
At low temperatures, on the other hand, cutoff effects
in the mesonic screening mass appear to 
be moderate, while at the same time the screening mass 
itself becomes small, $m/T\sim 1$ around $T \sim \Tc$,
indicating the ``vicinity'' of a second order transition. This 
suggests, in accordance with the discussion around \eq\nr{finiteV}, 
that {\em finite-volume effects} may be significant at 
{\em low temperatures}. 
For pure SU(3), this is clearly visible also in \fig\ref{fig:nf0}(left).
\ei
Hopefully these points serve to illustrate that there is  much
to learn from screening masses, and encourage them to be adopted
as a standard part of every finite temperature lattice study. 

\begin{figure}[tb]


\centerline{%
\epsfxsize=7.0cm\epsfbox{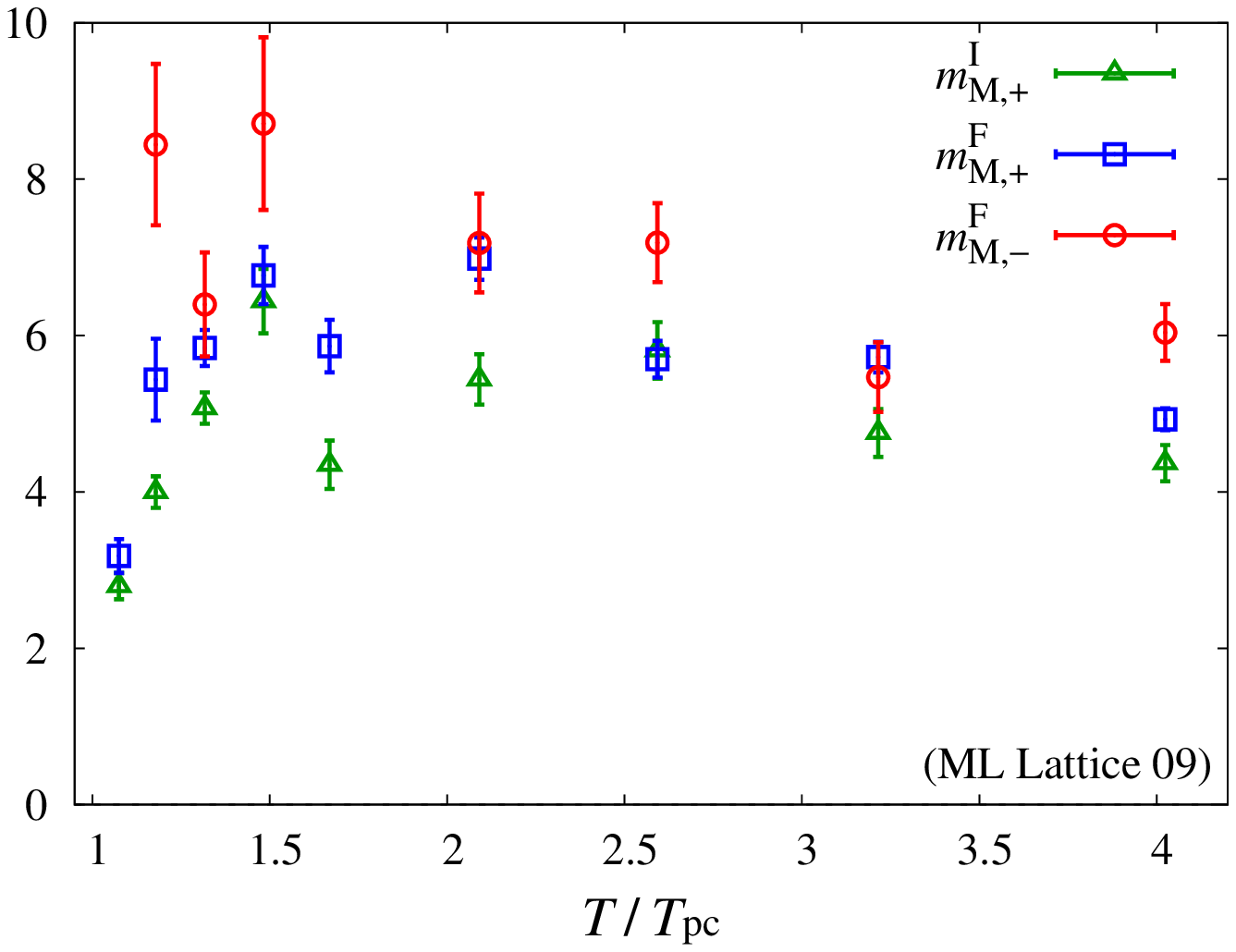}~~~%
\hspace*{0.5cm}%
\raise0.0mm\hbox{\epsfxsize=7.0cm\epsfbox{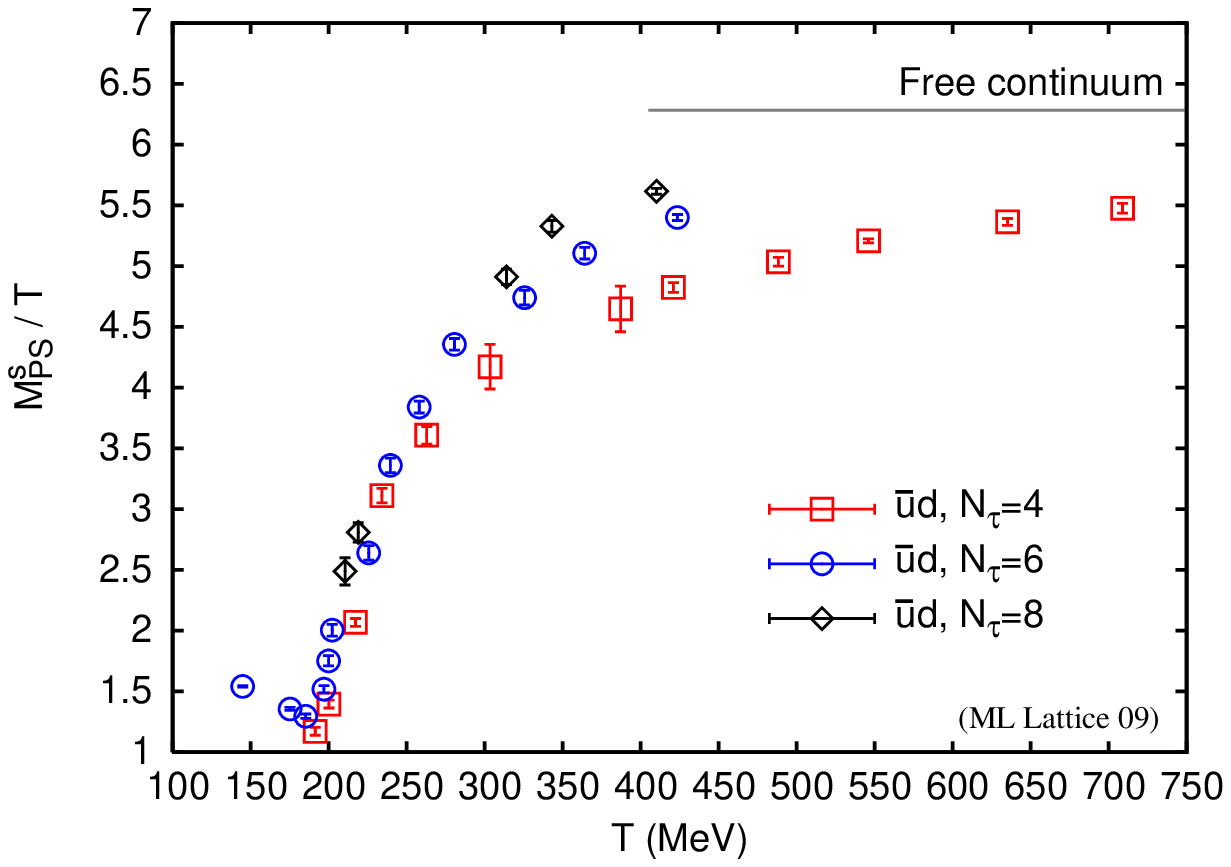}}~~~%
}


\caption[a]{\small 
Left: the smallest screening masses extracted from glueball 
like operators, with $\Nf = 2$ Wilson quarks in a $V=16^3\times 4$ 
volume 
(from ref.~\cite{whot2}).
Right:  the smallest screening mass extracted from a 
pseudoscalar mesonic operator, with $\Nf = 2 + 1$ p4fat3 
staggered quarks (from ref.~\cite{rbc2}).
} 
\la{fig:mT}
\end{figure}

%
\subsection{Energy-momentum correlators and sum rules}
\la{ss:sum}

I wish to end with a somewhat amusing recent episode, 
related to the determination of transport coefficients. 
As was already discussed in section~\ref{ss:1valence}, 
extracting a transport coefficient from a conserved current is
very difficult in general, because there may be a narrow transport 
peak in the corresponding spectral 
function, $\rho(\omega)$, around $\omega = 0$, 
and one would need to extract the height of the peak. 
If, however, analytic information is available on the shape of 
the spectral function, then sum rules relate 
integrals over $\rho$ to various thermodynamic quantities, 
which are easier to measure.

Now, a particular quantity considered in this spirit is the standard 
hydrodynamic transport coefficient known as bulk viscosity. The sum 
rule states that 
\be
 {S} = 
 \int_0^\infty \frac{{\rm d}\omega}{\omega} 
 \Bigl[\rho^\rmi{bulk}(\omega) - 
 \rho^\rmi{bulk}_\rmi{$T=0$}(\omega)
 \Bigr]
 \;, \la{sumrule}
\ee
where ${S}$ is a certain local thermodynamic observable. 
Setting aside the question (which I consider to be very
difficult) of whether the functional form of
$\rho^\rmi{bulk}(\omega) - 
 \rho^\rmi{bulk}_\rmi{$T=0$}(\omega)$
can be reasonably modelled, there has even been confusion about
how the left-hand side, ${S}$, looks like. 
In fact, two different answers have been proposed: 
one in refs.~\cite{ellis,kt}, and more recently another one in 
ref.~\cite{sr} (see ref.~\cite{sch2} for a related discussion). 
The two expressions are compared in \fig\ref{fig:sumrule}.

\begin{figure}[tb]


\centerline{%
\raise0.0mm\hbox{\epsfxsize=7.0cm\epsfbox{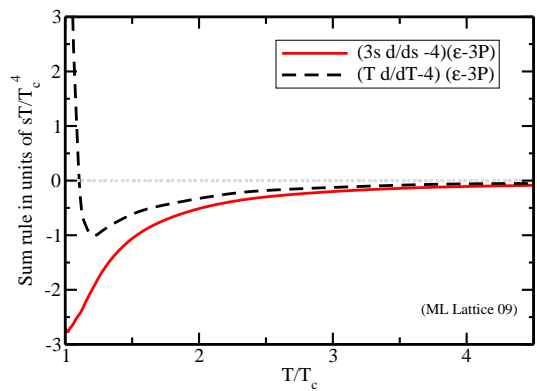}}~~~%
}


\caption[a]{\small 
A comparison of two different quantities that have been 
thought to be relevant for the bulk viscosity sum rule, \eq\nr{sumrule}
(from ref.~\cite{sr}). The lattice data is for pure SU(3)
and originates from ref.~\cite{boyd}.
} 
\la{fig:sumrule}
\end{figure}

The reason for the difference is in some sense subtle, 
though eventually perhaps also simple to understand~\cite{sr}. 
The older expression applies in Euclidean spacetime; 
it gives the value of a certain two-point function
at a fixed Matsubara frequency $\omega_n = 0$, 
extrapolated subsequently to vanishing spatial 
momentum. The other expression, in contrast, 
assumes first the limit of vanishing spatial momentum, 
while still keeping full information on all Matsubara
frequencies $\omega_n$, in order to allow for 
an analytic continuation to 
Minkowskian frequency $\omega$. 

Clearly, \fig\ref{fig:sumrule} demonstrates that 
it is important to have the correct left-hand side
in the sum rule of \eq\nr{sumrule}, if information
concerning the right-hand side is to be extracted. 

%
\section{Conclusions}

The main point which I wanted to 
illustrate in this talk is that, apart from 
the obvious need to continue numerical efforts in order to 
reach chiral, infinite-volume, and continuum limits with
controlled systematic errors for physical QCD,
there is also room and even 
need for various types of {\em theoretical} contributions to 
the understanding of strong interactions at high temperatures. 
This includes both analytic computations, examples of which were given 
in sections \ref{ss:1valence}, \ref{ss:finiteV} and \ref{ss:sum}, 
as well as numerical 
efforts in simplified theories, in which the systematic errors
can be controlled more easily than in full QCD; 
examples were mentioned in 
sections \ref{ss:qbq}, \ref{ss:nf0}, \ref{ss:nc4}. 
Hopefully more groups will join  also  these physically motivated 
``low-cost'' efforts in the coming years!

%
\section*{Acknowledgements}

I wish to acknowledge useful discussions and correspondence with 
S.~Bors\'anyi, 
B.~Bringoltz, 
F.~Bruckmann, 
S.~Datta, 
U.~Heller, 
L.~Levkova, 
Yu.~Maezawa, 
M.~Panero, 
A.~Rothkopf and \linebreak
G.~Schierholz, as well as financial support from the BMBF, under project
{\em Heavy Quarks as a Bridge between
     Heavy Ion Collisions and QCD}.


\end{document}